\documentclass[12pt,preprint]{aastex}

\def\kms{\relax \ifmmode {\ \rm km s}^{-1}\else \ km\ s$^{-1}$\fi}
 
\def\degree{\mbox{$^{\circ}$}}
\def\Mso{{M$_{\rm \odot}$}}

\def\cm3{${\rm cm}^{-3}~$}

\def\nii{[N~{\sc ii}]}

\def\hei{He {\sc i}}
\def\heii{He~{\sc ii}}
\def\oiii{[O~{\sc iii}]}

\def\ha{H$\alpha~$} 
\def\hb{H$\beta$}
 
\slugcomment{}

\shorttitle{CSPNs in the Large Magellanic Cloud}
\shortauthors{Villaver, et al.}
  
\begin{document}
  
\title{Post-AGB Evolution in the Large Magellanic Cloud. A Study of the 
Central Stars of Planetary Nebulae
\footnote{Based on observations made with the NASA/ESA Hubble Space Telescope,
obtained at the Space Telescope Science Institute, which is operated by the 
Association of Universities for Research in Astronomy, Inc., under NASA 
contract NAS 5--26555}}
   
\author{Eva Villaver}
\affil{Space Telescope Science Institute, 3700 San Martin Drive,
Baltimore, MD 21218, USA; villaver@stsci.edu}
\author{Letizia Stanghellini\altaffilmark{2}}
\affil{Space Telescope Science Institute; lstanghe@stsci.edu}
\author{Richard A. Shaw}
\affil{National Optical Astronomy Observatory, 950 N. Cherry Av.,
Tucson, AZ  85719, USA; shaw@noao.edu}

\altaffiltext{2}{Affiliated with the Hubble Space Telescope Space Department
  of ESA; on leave from INAF-Osservatorio Astronomico di Bologna} 
 
\begin{abstract}
We present medium- and broad-band Hubble Space Telescope ({\it HST})
photometry of a sample of 35 central stars (CSs) of Planetary Nebulae (PNs)
in the Large Magellanic Cloud (LMC). The observations were made with the
WFPC2 and STIS instruments on board the {\it HST}. By observing LMC objects, 
our sample is free of the distance uncertainty that is the dominate source 
of error in the determination of CS luminosities in Galactic PNs.  By
observing with the {\it HST}, we resolve the nebula and therefore 
we often detect the CSs unambiguously.  
We obtain core masses of 16 of the objects by comparing their positions 
on the HR diagram to theoretical evolutionary tracks, once we determine the
stellar effective temperature through Zanstra analysis.  
This sample of CS masses is the largest and most reliable set obtained in 
an extra-Galactic environment. We find an average mass of 0.65~\Mso,
though a few of the objects have very high mass. This average value is
consistent with the average mass of the white dwarf 
population in the Galaxy.  As the immediate precursors of white dwarfs, 
the study of the mass distribution of PN CSs should help to constrain the 
initial-to-final mass relation within environments of differing metallicity. 
Finally, by exploring the physical connections between the 
star and the nebula, we establish the importance of the study of PNs in the
LMC to constrain the energy input from the wind during the post-AGB phase.

\end{abstract}


\section{INTRODUCTION}
Central stars of planetary nebulae (CSPNs) are the result of the
evolution of stars in the approximate mass range 1--8 \Mso~that ascend 
the Asymptotic Giant Branch (AGB) after hydrogen has been exhausted 
and helium has been ignited in the core. 
During the AGB phase, low- and intermediate-mass stars
experience high mass loss rates that remove most of the stellar envelope,
leaving behind a stellar core that later on will ionize the previously
ejected envelope. The star then enters its evolution in the PN 
domain. During the PN phase the CS evolves at constant luminosity towards
higher effective temperatures, and later descends a white dwarf 
cooling track after the nuclear energy sources have been exhausted.

The upper initial mass limit for white dwarf production, according to 
stellar evolution theory, depends on the treatment of two poorly understood 
mechanisms: mass-loss and convection \citep{Blo:95,Her:00}. Therefore, an
observational determination of the initial--final mass relation, and 
therefore the minimum mass of type II Supernova progenitors, depends 
strongly on the measurements of white dwarfs masses.

White dwarfs are observed to posses a very narrow mass distribution which
peaks at $\sim$0.57~\Mso~and has a tail extending towards
larger masses \citep{Bsl:92,Fkb:97}. The CSPNs in the Galaxy are found
to peak around the same mass value \citep{Svmg:02}. However, the initial-to
final mass relation is expected to change slightly with the metallicity
\citep{Wei:87}, and because of the lower metallicity, the upper mass limit
of white dwarf progenitors is expected to be smaller in the Large
Magellanic Cloud (LMC) \citep{Uetal:99,Detal:99,Getal:00}. The mass
distribution of CSPNs in the LMC should reflect this fact as they are
the immediate progenitors of the white dwarfs population. Only four
masses of CSPNs in the LMC have been determined from direct
measurement of the stellar flux \citep{Detal:93,Betal:97}. The mass range of
3 of these CSs agrees with the range of values found in the Galaxy, and the
fourth one has a high mass progenitor. Thus, more mass determinations of
CSPNs in the LMC are needed to address the scientific problems described 
here.  

The determination of accurate masses of CSPNs is also important  in
order to solve a longstanding problem in PN formation: the likelihood that
the mass of the progenitor star determines the morphology of the PN it
hosts. Understanding the development of the simplest variety of shapes, from
round to bipolar, displayed by PNs is one of the most exciting challenges of
the late stellar evolution studies. On one hand, there is a large amount of 
observational evidence that shows fundamental differences in the physical 
and chemical properties between morphological classes. On the other hand,
there is still a wide debate on which of the proposed collimation mechanisms
operates in PNs, although numerical models are able to reproduce the 
overall  morphologies. While it seems clear that the mechanism is related 
to the stellar progenitor, it is to be determined whether the mass of the
progenitor, its magnetic field 
\citep{Pas:92,Glrf:99}, its rotation \citep{Cp:83,Glrf:99} the presence of a
companion star \citep{Ls:88}, or a sub-stellar object \citep{Ls:02} plays the
dominant role.

The importance of the progenitor mass in the development of the PN
morphology first suggested by \cite{Gre:71} has been corroborated from the
N and O chemical enrichment found in the bipolar and extremely axisymmetric 
morphological classes \citep{Pei:78,Tpp:97}. Bipolar PNs in the Galaxy are 
also found at a lower average distance from the Galactic plane than other 
morphological classes \citep{Cs:95, Metal:00, Svmg:02}, suggesting that 
they evolve from more massive progenitors. 
The correlations between the CS mass and the morphology for Galactic PN
samples has been explored by several authors
\citep{Scs:93,Amn:95,Cs:95,Gst:97,Svmg:02} who have found slightly
different mass distribution for the CSs of symmetric and axisymmetric
PN. However, determining CS masses for a statistically significant sample 
in the Galaxy is not an easy task. Typically, 
the evolutionary tracks of the CSs of different masses in the HR diagram
show very little variance with the CS luminosity, and 
since distances to Galactic PNs are very uncertain, so is the estimation
of their luminosities. Moreover, CSPNs are faint and the nebular 
continuum emission can completely mask the CS. 

{\it HST} offers a unique opportunity to study the CSPNs in the 
Magellanic Clouds, and to explore correlations with PN morphology 
and with PN physical conditions, with unprecedented accuracy, largely 
because their distances are independently known. 
In this paper we determine accurate masses of CSPNs for the largest 
sample of extra-galactic PNs ever studied, and we explore the connections 
among the fundamental properties of the stars (luminosity, temperature, 
and mass) and the physical properties of the host nebulae.

We present photometry of 35 CSPNs in the LMC obtained from the Cycle 8 
{\it HST} snapshot survey of LMC PNs using broad-band imaging with the 
Space Telescope Imaging Spectrograph (STIS), and from the Cycle 9 {\it HST} 
medium-band F547M (Str\"omgren $y$) images obtained with Wide Field 
Planetary Camera 2 (WFPC2).  The broad band images of the 29 PNs observed 
with STIS have been already published by \citet{Setal:01} (hereafter
Paper I), and the line intensities and nebular physical conditions obtained
by using {\it HST} STIS slitless spectroscopy by \citet{Setal:02} (hereafter
Paper II). In \S2 and \S3 we describe the observations and the photometric
calibration. The CS temperature and luminosity determinations  and their
distributions versus different nebular parameters are presented in \S4. The
results are presented in \S5 and discussed and summarized in \S6.

\section{OBSERVATIONS}
\subsection{The STIS Broad-Band Data}

The STIS observations of 29 of the targets presented in this paper are 
from the {\it HST} GO program 8271.  The observation log, observing 
configuration, target selection, acquisition, and a description of the 
basic calibration (through flat-fielding) can be found in Paper I. The
photometry of the CSs was measured from the STIS clear aperture mode 
images (50CCD). The 50CCD is an unvigneted aperture with a 
field of view of 52\arcsec$\times$ 52\arcsec~ and a focal plane scale of
0\arcsec.0507 pix$^{-1}$. In this setting no filter is used and the shape of
the bandpass is governed by the detector (which has a sensitivity from 
$\sim$2,000 to 10,300 \AA), and by the reflectivity of the optics. The 
central wavelength of the 50CCD is 5850\AA, and the bandpass is 4410 \AA. 
The FWHM of a PSF is close to 2 pixels at 5,000 \AA, and the 90\% 
encircled energy radius is 3 pixels \citep{Letal:01}.  The observations 
were made with the CCD detector using a gain of 1 $e^-$ per analog-to-digital 
converter unit (ADU). All the exposures were split into two equal 
components to facilitate cosmic-ray rejection. 

Table~1 gives in column (1) the object name, in column (2) the instrument 
and configuration used for the observation, in column (3) the total 
integration time, and column (4) gives whether or not the CS was detected 
in the images. 

\subsection{The WFPC2 F547M Data} 

We obtained images of 13 PNs with the WFPC2 instrument on {\it HST}
between April 2000 and May 2001. The WFPC2 observations were executed in 
GO program 8702, which aimed to recover CSs that were undetected 
in program 8271, owing to severe contamination from nebular continuum, and 
in GO program 6407 (P.I. Dopita), where only narrow-band images were 
available. Of the 13 targets observed in program 8702, seven are in common 
with program 8271 and the other six objects are from program 6407.

Two exposures were taken for each of two closely-spaced (dithered) 
pointings, with the object centered on the PC reference aperture.  The 
spatial scale is 0\arcsec.0455 pixel$^{-1}$. The observations were taken 
with the filter F547M at a gain of 7 $e^-$ ADU$^{-1}$. The medium-band 
F547M (Str\"omgren {\it y}) filter (centered at $\lambda$~5454~\AA~with a 
bandpass of 487~\AA) is a close match to the Johnson $V$ filter, though 
the bandpass is narrow enough to exclude the strongest nebular emission 
lines 
(\oiii~4959, 5007~\AA, \ha~6563~\AA, and \nii~6548+83~\AA).  However, 
nebular continuum emission is present in our images that originates 
mainly from the recombination of hydrogen.  In addition, F547M includes 
some contribution from weak emission emission lines such as \hei~5876~\AA, 
\heii~5411~\AA, and [\ion{Cl}{3}]~5527~\AA. 
Another possible source of nebular contamination is the `leakage' of the 
\oiii~5007~\AA~emission into the F547M filter pointed out by 
\cite{Retal:02}. The radial velocities of the LMC PNs we observed are between
220 and 285 \kms 
\citep{Metal:88}. The central wavelength for these objects will be displaced 
towards the red by about 4-5~\AA. This effect will increase the filter 
transmission in the \oiii~5007~\AA~line, which is usually one of the 
brightest in PNs, by about 0.4\%. So for our data, some emission from
the [\ion{O}{3}] 5007~\AA~line in the F547M filter is very likely.

The images were calibrated using the standard {\it HST} data pipeline (see
\citealt{Bag:02}). Duplicate exposures were combined, but with 
rejection of cosmic rays. Further rejection of cosmic rays and hot pixels 
was applied when the dithered images were aligned and co-added. 

\section{ANALYSIS}
\subsection{Photometric Technique}

The application of conventional photometric methods to CSPNs in the
Magellanic Clouds is currently only possible with the spatial resolution 
offered by {\it HST},
where the nebula is resolved and the separate the nebular and stellar 
contributions to the emission can be distinguished.  We have applied 
aperture photometry techniques to our data using the 
IRAF\footnote{IRAF is distributed by the National Optical Astronomy
Observatory, which is operated by the Association of Universities for
Research in Astronomy, Inc., under cooperative agreement with the National
Science Foundation.} 
{\bf phot} task.  Briefly, we measured the flux within a circular aperture 
centered on the star. The flux within this aperture also includes the 
nebular emission projected onto the star, for which we correct by 
subtracting the nebular flux in an annulus surrounding the aperture 
of the star. The nebular emission may be very inhomogeneous, so for each
object we evaluated the radial distribution of nebular flux in 
order to select the optimal aperture width and position.

The stellar aperture was chosen to have a radius of 2 and 3 pixels for 
the STIS and the WFPC2 images, respectively. Bigger apertures increase the
noise without increasing the signal and a smaller aperture is not advisable 
since the enclosed flux will depend on the position of the star within the 
pixel.  The fraction of the stellar PSF that falls outside the stellar 
aperture is taken into account with an aperture correction that is well 
determined from the instrument PSFs. 

For most of the objects in our sample, the contribution of the nebula can 
be accurately subtracted by using the median of the flux in an annulus with 
a width of 2 pixels adjacent to the stellar aperture. Strong variations 
around the median of the subtracted nebular flux are reflected in the 
standard deviation and, therefore, are propagated into the errors of the 
measured magnitudes. 
In those cases where the nebular emission decreases very sharply with the 
radius (e.g., for very compact PNs), or when the CS emission is faint 
compared to the nebula, an accurate value of the nebular flux for subtraction
could not be determined with this technique without very large errors in 
the photometry. In those cases we performed the photometry on an image 
where the two-dimensional nebular emission was removed.  We constructed a 
nebular emission image for this purpose by coadding the available 
monochromatic images taken from the STIS spectroscopy (see paper II), where 
the individual monochromatic images were weighted by the throughput of the 
50CCD bandpass for that wavelength.  We considered the \hb, \oiii~4959, 
5007~\AA,\ha, and \nii~6548 and 6584~\AA~ contribution.  We subtracted the 
resulting nebular image from the 50CCD image.  We then used the our annulus 
subtraction technique to eliminate any residual nebular contribution.  

To test the validity of the procedure we have applied the two methods
described above to four randomly selected PNs (SMP~4, SMP~10, SMP~27, and
SMP~80), that is, aperture photometry on a nebular subtracted
image and aperture photometry when the nebula has not been previously
subtracted. In both cases the emission of the nebula, or the residuals from
the nebular subtraction, are determined in an annulus surrounding the stellar
aperture. The differences in the magnitudes measured in the cases where we
tested both methods are at the 0.001 magnitude level, which is smaller than
the 
errors in the magnitudes. We believe that for most of the objects it is
not necessary to subtract a nebular image before performing the 
aperture photometry.  But in those cases for which either 
the CS is detected at a very low level above the nebular emission or the 
nebula is very compact, we were able to reduce the errors greatly by 
subtracting the nebula prior to performing the photometry.

When the CS is not detected (i.e., no stellar PSF appears above the nebular 
level), we computed a lower limit to the CS magnitude by measuring the flux 
inside a stellar aperture at the geometric center of the nebula (i.e., the 
most likely position of the CS). The nebular background flux was measured 
in an adjacent aperture and then subtracted. The lower limit to the stellar 
magnitude is the measured magnitude, minus the error in magnitudes. 
Obviously, the lower limits for the magnitude are very uncertain since 
they depend on the flux measured within an aperture that may or may not 
actually contain the CS. Moreover, it is very
difficult to set the nebular level when the CS is not visible, since it
depends strongly on a very uncertain position, which could contain the CS. 

\subsection{Photometric Calibration}
\subsubsection{The STIS Broad-band Data}

We have transformed our net, instrumental count rate to magnitudes measured 
in the STMAG\footnote{The STMAG is the Space Telescope magnitude
system, based on a spectrum  with constant flux per unit wavelength.}
system by using the zero-point calibration given by
\cite{Betal:02}.  (The zero-point used was 26.518.) The STIS charge transfer
efficiency (CTE) has been characterized by \cite{Ggk:99} and the effect on
the magnitudes has been shown to be below 0.01 mags \citep{Retal:00}, except
for very faint stars on the edge of the CCD, which was never our case. 
Therefore, we have ignored the CTE correction for the STIS data since
it is negligible for our purposes. In this observing mode, the 
image distortions can be neglected as well since they are less than a pixel
across the whole detector. The aperture correction applied to the magnitudes
measured in a radial aperture of 2 pixels is 0.517 dex, based upon the curve
of encircled energy derived by \citealt{Betal:02} for stars near the field
center.  

\subsubsection{The WFPC2 F547M Data}

The zero-point calibration to the STMAG system for the filter
and observation configuration of the data taken with the WFPC2 was taken
from \cite{Dol:00} (we used a zero-point of 21.544). We have applied the CTE,
geometrical distortion, and aperture corrections to this data. The CTE
correction, which depends on the position on the chip, target brightness, 
background, date, and observing mode, was determined for our data following 
the prescriptions of \cite{Dol:00}. The geometrical distortion in
the WFPC2 field, which is removed during calibration during flat-fielding,
causes pixels to have different effective areas as a function of position.
It does not affect surface photometry but it affects point source
photometry. Therefore, we have applied a correction for geometrical
distortion depending of the position of the star on the
CCD by using a geometric correction image. We have determined the
offset between our aperture (3 pixels) and the nominal aperture used for the
calibration (0\arcsec.5 in radius, which correspond to 11 pixels for the PC
camera) by selecting isolated,
bright stars on each field and averaging the difference between the 3 pixel
and the 11 pixel apertures.  Usually, we have averaged the values for five 
stars on each field. No contaminant correction was applied 
since it is significant only for UV observations.
In order to check the calibration of our STMAG instrumental magnitudes we
performed point-spread function (PSF) photometry on each field with the
{\bf HSTphot} package \citep{Dol:00b}. Then we verified that the magnitudes 
of selected stars on the field measured with {\bf HSTphot} agreed with our 
aperture photometry measurements after all the corrections were performed.  

\subsection{The Extinction Correction}

To derive the stellar extinction correction we used the nebular Balmer 
decrement. We adopted from Paper II the extinction constants for all objects
except  
for SMP~50, SMP~52, SMP~56, and SMP~63 (which were taken from 
\citealt{MED}), and SMP~33 and SMP~42 (taken from \citealt{MED2}). 
The conversion from the nebular extinction constant, {\it c} (the logarithm 
of the total extinction at \hb), to the color excess, E$_{B-V}$, 
requires some caution \citep{Kl:85}. The approximate relationship between 
{\rm c} and E$_{B-V}$ depends upon the spectral energy distribution (SED) 
of the target in question. We have used the approximate relation 
$c=1.41$ E$_{B-V}$. \cite{Kl:85} found that the ratio of {\it c} to 
E$_{B-V}$ shows little variation with the stellar  temperature, but increases 
with the amount of extinction.  Since the amount of extinction measured in
our objects is typically small, we are confident about the assumption of a
constant value for the relation between {\it c} and E$_{B-V}$. Adopting a
different relation is only meaningful in the 
cases where $c \ge 0.2$, which leads to E$_{B-V}$ values that change
in the second decimal place (i.e., of the order of our photometric errors). 

The reliability of our method to determine the stellar extinction assumes
that the extinction does not vary across the nebula due to internal 
absorption by dust.  In Paper II no significant variations of the \hb/\ha 
ratio were found for heavily reddened objects on spatial scales of 
$\sim$0.04 {\rm pc} which give us confidence on the use of {\it c} to 
derive the amount of stellar extinction.

In the wavelength range under consideration, the LMC extinction law is very
similar to the Galactic extinction law \citep{How:83}.  Thus, in order to
convert E$_{B-V}$ to total absorption in the $V$ band, we used the 
interstellar extinction law of \cite{Sm:79}, and assumed that R$_{V} = 3.1$. 
The extinction in magnitudes (A$_{V}$) is then A$_{V} = 2.2 c$.

\subsection{The Transformation to Standard $V$ Magnitudes}

The filters in {\it HST} instruments do not match perfectly the bandpasses 
of standard photometric systems, such as Johnson-Cousins {\it UBVRI}, so 
the transformation from instrumental magnitudes to a standard system depends 
on the SED of the object observed. For the WFPC2 data the transformation is 
straightforward because the F547M filter is a close match to the Johnson 
$V$ filter (see \citealt{Bir:02}). Our WFPC2 F547M magnitudes have been
transformed to the standard V magnitudes following the prescriptions of
\cite{Hetal:95}. The  color (V-I) needed to apply the transformation has been
derived via synthetic photometry with IRAF/STSDAS {\bf synphot} using a
blackbody spectrum to represent  the SED of the CSs. The color dependence 
with temperature and reddening has been determined using as input a range 
of blackbody temperatures and E$_{B-V}$ in the parameter range of our 
CSPNs. For our purposes, it seems that a blackbody is as good an 
approximation as 
any model atmosphere to represent the SED of CSPNs \citep{Gkm:91}. We find 
that the transformation from WFPC2 F547M to standard $V$ magnitudes is
rather insensitive to changes in the stellar temperature and reddening 
within the range of values of our sources.  The median of the transformation 
obtained by using a ($V-I$) color range derived for CS effective 
temperatures between 30,000 and 300,000 {\rm K}\footnote{The lower 
temperature limit is set to provide enough ionizing photons and the upper 
limit is taken from the CS evolutionary tracks of \citealt{Vw:94}} is -0.013
magnitudes, with a standard deviation of $\sigma = 0.002$.
By using a CS  effective temperature of 50,000 {\rm K} and
allowing E$_{B-V}$ to change between 0.1 and 1, we obtain a transformation
with a median of .001 magnitudes and a $\sigma$ of 0.004. We have converted
the WFPC2 magnitudes to standard $V$ by using the median of the transformation
obtained for  effective temperatures between 30,000 and 300,000 {\rm K}. We
have estimated the error of the transformation to be the 
quadratic sum of the $\sigma$ 
obtained for  the range of effective temperatures and the $\sigma$ obtained
for the E$_{B-V}$ range considered. 

The STIS 50CCD bandpass is very broad and its response curve is far from 
that of the standard $V$ filter. Moreover, the transformation from 50CCD 
to $V$ magnitudes has not been published. 
Therefore, obtaining standard $V$ magnitudes from our 50CCD data requires 
considerable care. As a first step we used the {\bf synphot} package in 
STSDAS to explore the dependence of the transformation with both the 
E$_{B-V}$ and the CS temperature. We explored the CS temperature range
30,000-300,000 {\rm K}, and the E$_{B-V}$ in the range  of our data. We find
that the correction is strongly dependent on the extinction and, to a lesser
extent, on the temperature of the star.  Therefore, we have determined the
transformations of the 50CCD to V individually for  each object by
determining the median of the the $V-$50CCD colors for blackbodies between
30,000 and 300,000 {\rm K} and using the E$_{B-V}$ value of each source. The
$\sigma$ of the transformation for each object (given for the correction in
the range of temperatures) has been added to the error of the magnitude. The
highest standard deviation we get is 0.05 magnitudes and the highest values
of the 
correction is 0.308 magnitudes (for J41, SMP~59, SMP~93, and SMP~102 with 
E$_{B-V} = 0$), though most of the corrections near 0.1 magnitudes.

In Table~2, we give the results of the photometry. Column (1) gives the PN
name (according to the SMP nomenclature when available); column (2) gives 
the $V$ magnitude or its lower limit, as well as the associated errors.  The 
error value includes the random (photon noise, read noise), systematic 
(CS flux, sky), and the errors in the calibration. In those cases where 
the data were saturated we note that circumstance in the table with the 
measurement of the magnitude. The magnitudes derived from saturated data 
have not been used for the analysis in the rest of the paper. Unless noted
otherwise the magnitudes obtained from the STIS data are given. The color
excesses used to correct for extinction are listed in column (3). 

\section{The Determination of the CS Effective Temperature}

The temperatures of the CSs were computed using the Zanstra method 
\citep{Zan:31}. The method, fully developed by \cite{Hs:66}, and extensively
used in the literature (i.e. \citealt{Kal:83}), derives the total ionizing
flux of the star by comparing the flux of a nebular recombination line 
of hydrogen or helium to the stellar continuum flux in the $V$ band.  The 
method assumes a particular choice of stellar spectral energy distribution, 
which from now onwards we consider to be a blackbody. 
The Zanstra method also assumes that all the photons above the Lyman limit 
of H or He$^+$ are absorbed within the nebula and that each recombination 
results in a Balmer-series photon. Therefore, when the \heii~4686 \AA~line 
flux is available the Zanstra method gives two values of the stellar 
temperature. 

The data needed for the temperature calculation were taken from Paper II 
(\hb~fluxes, nebular radii, and extinction constants) except for SMP~33, 
SMP~42, SMP~50, SMP~52, SMP~56, and SMP~63 where the \hb~fluxes were taken 
from \cite{Mdm:88}. The \heii~4686~\AA~line fluxes were taken from 
\cite{BL}, \cite{MED2}, \cite{VDM}, \cite{JK93}, and \cite{MBC}. The line 
intensities given in some of these references were corrected for extinction, 
so we have uncorrected these fluxes using the extinction constants given 
in the original references and the average Galactic reddening curve of 
\cite{Sm:79}. In order to assure the best results we have been very 
conservative with the errors in the fluxes quoted by the  references. 
We have supplemented the above with fluxes from our unpublished, ground-based
for SMP~33, SMP~56, SMP~100, SMP~102, SMP~34 and SMP~80 
\citep{Setal:03, Petal:03}. In Table~2, column (4) we list the \heii~ 
4686 \AA~line intensity (and error) relative to \hb~=~100, not corrected 
for extinction; in column (5) we list the reference code for the
\heii~fluxes.  
 
\subsection{Bolometric Corrections and LMC Distance Estimates}

We computed bolometric luminosities for the CSs in our sample. 
The bolometric correction (BC) dependence with T$_{eff}$ was taken from
\cite{Vgs:96} which was derived for Galactic O-type and early B-type
stars. We use  this relation since the dependence of the BC on $\log$~{\it g}
was found to be extremely weak. \cite{Flo:96} also found that all luminosity
classes appear to follow a unique BC-T$_{eff}$ relation.  The BCs have been
computed by using the 
\heii~Zanstra temperature when available, otherwise the H~{\sc i} Zanstra
T$_{eff}$ was used.  Temperatures from T$_Z$(\heii) are the most reliable 
because of the likely 
optical thickness of most PNs to \heii~ionizing photons. T$_Z$(H) can be
reliable for PNs with sufficient optical depth; the problem is to determine 
which PNs are optically thick to hydrogen ionizing radiation. The derived 
BCs agree with the empirical values given by \cite{Cod:76}.

In order to compute the CS luminosities we adopted a distance to the LMC 
of 50.6 {\rm Kpc}, and an absolute bolometric magnitude for the Sun of 
M$_{bol,\odot}$= 4.75 mag \citep{All:76}. We 
estimated the error introduced in the derivation of the luminosity due
to the distance variation caused by the depth of the LMC.  The LMC can be 
considered as a flattened disk with a tilt of the LMC plane to the plane of
the sky of 34.7\degree \citep{Vdm:01}. \cite{Fio:83} derived a scale height
of 500 {\rm pc} for an old disk population. The scale height of young objects
is between 100 to 300 {\rm pc} \citep{Fea:89}. From the three dimensional
structure of the galaxy, we have a spread in the distance modulus of 0.03
which we propagate into the error of the absolute magnitudes and
luminosities. We have not taken into account the errors in the distance to
the LMC, since it will affect all the objects in the same way.

We give our resulting temperatures and luminosities in Table~3.  In 
column (1) we give the PN name; in columns (2) and (3) we give the 
effective temperatures (in units of 10$^3$ {\rm K}) derived from the
Zanstra method for the \heii~and for the hydrogen recombination lines, 
respectively. The two luminosity determinations given by the Zanstra method
are given in columns (4) and (5) respectively.  Where the CS was not 
detected we give the upper limit of the luminosity. The visual absolute 
magnitude and the stellar luminosity (derived from the BC) are listed in 
columns (6) and (7), respectively. The BCs and their errors, computed by
propagating the errors in the determination of T$_{eff}$, are given in 
column (8). All the values are listed with their respective errors.

\section{RESULTS}

In the following discussion we use the morphological classification of the 
nebulae provided in Paper II. PNs are classified as Round, Elliptical, 
Bipolar, Bipolar Core, and Point-symmetric according to their morphology 
in the \oiii $\lambda$ 5007 line. In those cases where we have two
measurements of the  
magnitude, we have selected one of them according to the following
criteria: for SMP~9, SMP~16, SMP~46 and SMP~53 we have used the STIS lower
limits on magnitudes since they are deeper observations; for SMP~19 and
SMP~30  
we used the STIS data because the CSs are detected with this instrument; 
and in the case of SMP~78 we used the WFPC2 data because the STIS data are 
saturated. All saturated and unresolved objects have been excluded from the 
following analysis. 

\subsection{The Effective Temperatures}

In Figure~1 we plot the Zanstra ratio T$_Z$(\heii)/T$_Z$(H) vs. T$_Z$(\heii), 
but only for those objects for which the CS was clearly detected (i.e., no 
limiting values were used), and where T$_Z$(\heii) and T$_Z$(H) are the
Zanstra temperatures derived from the \heii~4686\AA~ and the
\hb~recombination lines, respectively.  
The difference between the two temperature determinations is a 
well known effect: the ``Zanstra discrepancy'' \citep{Kal:83, Kj:89, Gp:88}. 
The Zanstra discrepancy has been studied by several authors; the optical 
thickness in the nebula to the H and He$^+$ ionizing radiation is the 
principal reason cited for T$_Z$(\heii) often exceeding T$_Z$(H) 
\citep{Kj:89,St:86,St:90, Gv:00}. 
We find that the Zanstra ratio approaches the unity for higher effective 
temperatures, in agreement with the previous results cited above. SMP~10 
and SMP~56 (the only filled round point and the only open circle, 
respectively) are the only objects which have both a small Zanstra ratio 
and a low effective temperature. Both objects have a small \heii~flux which
can indicate either that the nebula is optically thick to both hydrogen and
helium radiation or that the \heii~Zanstra calculation is rather uncertain.
SMP~10 is also the only point-symmetric object in this sample and SMP~56 the
only round PN in the plot. It might be significant that both fall off of
the general trend. Although the number of objects is too small to be 
conclusive, we do not find any morphological segregation of the Zanstra 
discrepancy in Figure~1, with the exceptions of SMP~10 and SMP~56. 

\cite{Vmg:02} found that the transition from an optically thick to an 
optically thin nebula depends on the initial mass of the star: the higher
the initial mass, the higher the effective temperature at which the nebula 
becomes optically thin. If the Zanstra discrepancy is due only to the optical
thickness in the H ionizing radiation, and since the Zanstra ratio approaches
the unity for higher effective temperatures, then, according to the results
of Villaver et al. (2002b), it is very likely that the objects with the higher
Zanstra discrepancy have low mass progenitors. We will return to this point
in the discussion.

\subsection{Luminosities}

In the adopted Zanstra method, the stellar temperature is computed by
determining the 
point where a parameterization of the stellar luminosity with temperature
(based on a blackbody assumption and on the measured visual magnitude and
extinction), equals the parameterization for the luminosity of the nebula in
two different recombination lines. Thus, the Zanstra method gives as a
by-product two determinations of the stellar luminosity: from the hydrogen
and the \heii~temperatures. The Zanstra method does not use
any empirical bolometric correction to convert the magnitude in the
visual band to the bolometric magnitude, but is instead a function of the
stellar temperature, which itself is based on a blackbody assumption.

The CSPNs in the LMC are free of the distance uncertainty that
dominates the determination of CSs luminosities for Galactic PNs. However,
another problem remains: the measurement of the T$_{eff}$ that influences
the luminosity determination through the BC. We have the stellar luminosities
derived from the Zanstra method (L$_Z$) and those derived from the observed
magnitudes using the BCs (L$_*$). Although the T$_{eff}$ plays a role in 
both determinations, they are not completely independent and we can compare 
them to check the consistency of luminosity determinations based on these 
two approaches.

In Figure~2 we show the $\log$~L$_*$/L$_{\odot}$ versus the
$\log$~L$_Z$/L$_{\odot}$ derived from \heii~and  H~{\sc i} Zanstra analysis
(left and right panels, respectively). We are comparing the two luminosity
determinations in a self-consistent way, and the luminosities derived by using
the \heii~Zanstra temperature to determine the BC, are compared with those
luminosities derived from the \heii~Zanstra analysis (left panel). The same 
comparison is valid for the luminosities derived from hydrogen Zanstra 
temperatures. These are plotted in the left panel of Figure~2, but only 
for those CSs for which the \heii~Zanstra temperature was not available. 
Figure~2 shows that the points lie very close to the 1:1 relation.  

The two samples of CSs displayed in the left and right panels of Fig.~2 show
a quite different luminosity range. The CSs displayed in left panel of Fig~2,
those for which we have a  \heii~Zanstra temperature, appear to be more
luminous than the CSPNs with hydrogen Zanstra temperatures (right
panel). T$_Z$(\heii) is thought to represent more closely the effective
temperature of the star, and it is rather well established that T$_Z$(H) is
probably underestimating the CS temperature for optically thin objects. CSs
with higher effective temperature will have bigger BCs and therefore more
likely higher luminosities, and thus if the T$_Z$(H) is underestimated is
likely  that the CSs luminosities will be underestimated too. However, we do
not think that the lower luminosity range of the CSs in the left panel of
Figure~2 is due to this effect. The absence of 4686~\heii~line emission in
these nebulae suggests that the CSs have a low temperature, which is indeed
the case (i.e., the CSs are not hot enough to ionize
\heii). Therefore, we are more inclined to think that most of the CSs plotted
in the right panel of Fig.~2 are intrinsically low luminous stars. 
It should be mentioned that \heii~4686 fluxes of only 0.005 of
\hb~can yield to significantly higher T$_Z$(\heii) than T$_Z$(H) Zanstra
effective temperatures. Rarely 
\heii~4686 line fluxes are available at this accuracy thus the 
distribution in the log-L, log-T plane could change significantly with 
higher quality spectra. 
 
Figure~3 shows L$_*$, T$_Z$, and the nebular radius plotted (from top to
bottom) versus the relative difference between L$_*$ and L$_Z$ on a linear
scale. The  left and right panels represent the relative differences in the
luminosity when derived from the \heii~and H~{\sc i} Zanstra analysis,
respectively. In the left panels we plot only those objects for which a
\heii~Zanstra temperature is available. In the right panels we plot the
CSs with only hydrogen Zanstra temperatures, i.e., only the CSs for which 
the \heii~4686 \AA~flux relative to \hb~is zero or not available. 
We find that all the L$_Z$(\heii) are within  $\sim$30 \% of the L$_*$. We do
not find any systematic differences with L$_*$, T$_Z$(\heii), nor with the
nebular radius or morphology. The L$_Z$(H) and L$_*$ agree to within
20\%, with the exception of two objects (SMP~63 and SMP~65) which are within
40\%. SMP~31 is not shown in the plots because of the high
(L$_*$-L$_Z$[H])/L$_*$ $\sim$150\%. It
could be that there are systematics in the (L$_*$-L$_Z$[H])/L$_*$
difference, but this is not conclusive because of the small number of
data points. 
As mentioned before, the dependency of the BC with T$_{eff}$ is based on an
empirical approach, while in the Zanstra analysis the relation between the
luminosity and T$_{eff}$ is based on a blackbody assumption. We find that
both determinations provide similar values (within $\sim$30\%) of the CS
luminosity.  

\subsubsection{Luminosity-Nebular Radius and Surface Brightness relations}

The relation between CSs and the nebulae is explored in Figure~4 where we
show the $\log$~L$_*$/L$_\odot$ versus the  nebular photometric radius (taken
from Paper II). The photometric radius and the CS luminosity should be good
indicators of the evolutionary status of the nebula and the CS,
respectively. On one hand, the evolution of the nebular radius is a gas
dynamics problem which depends, among other things, on the energy that the
stellar wind is injecting into the gas, which is a function of the evolution 
of the stellar luminosity and the core mass. On the other hand, the stellar 
luminosity, after a constant phase for hydrogen-burners, decreases during 
the evolution at a rate that depends mainly on the core mass.  Thus, it 
is expected that both quantities decrease with time, and that they do not 
evolve independently; the evolution of the radius must be related with 
that of the luminosity of the CS.
 
We find in Figure~4 a tendency of higher radius for lower stellar
luminosities. There is an apparent segregation in Figure~4 of smaller nebular
radius for symmetric PNs for which a selection effect towards the detection
of younger symmetric PNs cannot be ruled out. \cite{Vmg:02} studied the PN
formation for a range of progenitor 
masses and followed the gas structure that resulted from the AGB evolution
\citep{Vgm:02}. To have an indication of how the nebular evolution relates to
the evolution of the CS we have super-imposed on Fig.~4 an interpolation of
the nebular radius evolution with the CS luminosity for different progenitors
from the numerical simulations of \cite{Vmg:02}.

A qualitative comparison with the models shows that 
high luminosity objects with large radius may have a low mass progenitor
star because of the fast evolution of the CS luminosity for high mass
progenitors. The nebular radius does not evolve as fast as the luminosity,
although the amount of energy injected by the wind is higher. 

A quantitative comparison with the models shown is not possible, since the
core 
mass, stellar luminosity, stellar wind history and gas dynamical evolution
are dependent on the metal content of the gas and star.  A different
metallicity will change the efficiency of the wind-driven mechanism and the
cooling of the gas. Work is in progress on numerical simulations of PN
formation that reflect the metallicity of the clouds. We would like to point
out here that, once the models are performed for the 
LMC metallicity, a plot like the one shown in Figure~4 will be very useful in
order to constrain the wind energy injected during the post-AGB phase by
the star. 

Figure~5 shows two direct observational
quantities in  the LMC: the  absolute visual magnitude and the nebular radius
(in logarithmic scale). There is a strong correlation between the absolute
visual magnitude and the nebular radius. An evolutionary effect is a very
likely explanation for this correlation since as the CS fades, the radius
becomes bigger. There is also a strong correlation with nebular morphology,
at least for symmetric vs. asymmetric types. Shaw et al. (2001) noted the
tendancy of Round nebulae to have  
systematically lower expansion velocities, and they may therefore be 
older than their relative sizes would suggest. However, their small size 
and visually bright CSs might imply that they are young. Thus, using 
the nebular radius and visual magnitude as the sole indicators of 
their evolutionary state may be too simplistic.  We shall address 
the complicated interpretation of nebular kinematics in this context 
in a future paper. 

In Figure~6 we plot the CS luminosity versus the nebular surface brightness 
in the \hb~emission line (SB$_{H\beta}$, defined as the integrated line 
flux divided by the nebular area $\pi$~R$_{phot}^2$). The luminosity 
gradient steepens as SB$_{H\beta}$ declines, as expected by the common 
evolution of nebulae and stars. We do not find any
low SB object with high luminosity, nor any object with low CS luminosity
and high SB, with the exception of SMP~59 (the square point in the right 
upper corner of the plots). The relation is shown in the left panel for the
\oiii~line (the \oiii~SB has been taken from Paper II). The fact that the
relation is the same, although with higher dispersion (which might be due to
variations in oxygen abundance), for a collisionally excited (rather than 
recombination) emission line strengthens the point that a fundamental 
physical process, related to nebular evolution, must underlie the cause. 
 
In Figure~7 we have added to the luminosity versus SB$_{H\beta}$ plot those 
points for which we have only a lower limit to the magnitudes (identified 
with an arrow) and we have surrounded each point with a circle of size 
proportional
to the nebular radius. Low SB objects are always located towards the
position of low luminosity CSs and have a larger radius. The potential use of
this kind of plot as an indicator of the nebular age is very clear. We find 
that it offers a better diagnostic than the HR diagram for showing 
the evolutionary status of the nebula, confirming that an evolutionary effect
must underlie the SB--nebular radius relations found in Papers I and II.

\subsection{Stellar Distribution on the Log L-Log T Plane}

Figure~8 shows the distribution on the Log L-Log T plane of the CSs of
those PNs in our sample for which the CS was detected. The evolutionary
tracks have been taken from \cite{Vw:94} for stars with LMC metallicity. The
post-AGB evolution of a CSPN depends on its previous AGB evolution and on
the phase of the thermal-pulse cycle on which the star leaves the AGB 
\citep{Sch:83, Vw:94,Blo:95}. Whether the star leaves the AGB when 
helium-shell or hydrogen-shell burning is dominant determines the 
He-burning or H-burning nature of the subsequent post-AGB track. The lower 
mass models presented by \cite{Vw:94} are more efficient at producing 
He-burning post-AGB tracks, which they argue is a natural 
consequence of the mass-loss behavior during the AGB phase.  The mass-loss
rates on the AGB were artificially enhanced or diminished to control the
point of departure from the AGB in the He-burning tracks models for initial
masses  1.5 and 2 \Mso. It is important to note that the mechanism that
controls  the departure of the star from the AGB is unknown and therefore
artificially defined in the stellar evolutionary models.

We do not find the tendency reported by \cite{Detal:96} of size evolution
along the evolutionary tracks on the HR diagram. We do find that the 
luminosity vs. SB diagram is a better diagnostic of the evolutionary
status of the nebula. It is important to note that the masses derived by 
\cite{Detal:96} and \cite{Vetal:98} are based 
photoionization modeling of the optical spectrophotometry of individual
nebulae. The conclusions about the helium and hydrogen-burning
nature of the progenitors are based on a comparison between the dynamical
ages derived from observations of the nebula and a determination of the 
theoretical time-scale. The theoretical time-scale was obtained through 
an empirical fit to the expansion velocity as a function of the position 
on the HR diagram and theoretical evolutionary tracks of the CS. The 
physical relationship between the expansion velocity of the nebula and 
the CS luminosity is a problem that requires numerical simulations to be 
solved.

The same kind of distribution on the HR diagram that \cite{Sk:89} 
called ``Zanstra's wall,'' that is, the apparent high number of CSs with
4.9~$\le$ log T$_{Teff}$~$\le$5.1, can be seen in Figure~8, although the
number of objects is very small. As pointed out by
\cite{Sk:89}, it is very possible that the nebula becomes optically thin to
\heii~radiation above a certain stellar temperature in which case the 
Zanstra method will be providing only lower limits of the temperature for the
hottest CSs. 

We have derived the core masses based on the locations of the post-AGB tracks
at hand, and the interpolations between them. We have not derived core masses
for the four points that lie below the tracks. 
Only hydrogen Zanstra temperatures could be computed for those objects, 
and therefore their temperatures are very likely underestimated.  If their 
temperatures were higher, those points would move towards the upper left 
of the HR diagram (the BC will increase and therefore so will the derived 
luminosity), where the theoretical stellar evolutionary tracks predict 
the evolution of most CSPNs. Another, albeit very unlikely possibility 
is that those objects are the result of the evolution of a
progenitor less massive than the lower mass progenitor available in the
models of \cite{Vw:94}, e.g., 0.9 \Mso.

The core masses and the morphological classification of the nebulae are
summarized in Table~4. We do not find any correlation between the mass of the
CS and the morphology of the nebula. The mean and the median of 
the mass distribution are respectively 0.65 and 0.64 \Mso.
The number of objects is small, and therefore the distribution of the
objects in the HR diagram has limited statistical value. However, this is the
first time ever that CSPN masses have been derived without the distance bias
affecting the Galactic PNs, so these averages are extremely valuable. 

\section{SUMMARY AND DISCUSSION} 

The distribution of CSPNs in the HR diagram is rather uncertain for 
Galactic PNs, mainly for the lack of reliable distances. Because LMC PNs 
are seldom resolved from the ground, most of the previous determinations 
of CSPN masses are highly model-dependent, i.e., they are based on CS
luminosities derived  
solely from nebular fluxes \citep{Hlb:89, MBC, Kj:90, Kj:91, Detal:96, 
Detal:97, Vetal:98}. Although {\it IUE} spectra were employed in the work 
by \cite{Aetal:87} the CSs were not detected and therefore photoionization 
modeling was used to determine the CS parameters. In all, only four CS 
masses were previously determined from direct measurement of the stellar 
flux, and then only from the UV spectrum of the CS \citep{Detal:93, 
Betal:97}. \cite{Detal:93} estimated the mass of the CS 
of SMP~83 to be $\sim$1~\Mso, a extremely massive object which has been
classified as a Wolf-Rayet nuclei by \cite{Petal:95}. The CSs masses
determined by \cite{Betal:97} are in the range 0.62-0.68~\Mso.

We have performed photometry and derived luminosities for a sample of CSs in
the LMC. By observing LMC objects for which the distance is well known, we 
greatly reduce the uncertainty in the vertical axis of the HR diagram; 
by directly measuring the CS continuua with {\it HST} we eliminate the 
dependency on photoionization models in the determination of the stellar 
flux. Although significant uncertainty remains in the determination 
of effective temperatures, we have adopted a very conservative approach 
by using for most of the cases only effective temperature derived from the 
\heii~Zanstra analysis, which are considered to be the more reliable. 

The Zanstra discrepancy has been successfully explained in the past based 
on an optical depth effect \citep{Kj:89,Gv:00}.  Villaver et al. (2002b) 
showed an increased likelihood that PNs become optically thin at a lower
effective temperature for lower mass progenitors. We find that most of the  
objects with high Zanstra discrepancy indeed have lower progenitor masses, 
confirming the optical depth effect as the 
main reason to explain the Zanstra discrepancy. However, two of our targets
with high Zanstra discrepancies (SMP~34 and SMP~50) are among the most
massive and youngest of the sample. It is very unlikely that these two
nebulae are optically thin to hydrogen 
radiation, so some other effect e.g., an excess of photons above the
\heii~ionization threshold if the stars had pure H stellar atmospheres, 
must be invoked to explain the Zanstra discrepancy in these two objects. 

We find an average mass for our sample of 0.65 \Mso, which is close to 
(although slightly higher than) the average mass of white dwarfs in the 
Galaxy \citep{Fkb:97,Bsl:92}. \cite{Fkb:97} pointed out that the average 
mass of white dwarfs should be used with caution as it depends on the 
underlying distribution of masses, which is a function of the temperature 
range covered by the sample. 
The total number of objects in our sample is very small so it would 
be premature plot histograms of our distribution and assign great 
significance to the peak values of the sample.  Work on the data of the 
{\it HST} SNAP program 9077 (where $\sim$ 60 LMC objects were observed) 
is in progress and will help us to address this issue. 

The mass-loss during the AGB phase is expected to be affected by the
metallicity, and the relation between the mass of a white dwarf and that of
its progenitor on the main sequence tell us the complete, integrated mass
loss through the evolution. If mass-loss is reduced in the LMC with respect
to the Galaxy because of the lower metallicity, and assuming that other
selection effects are not operating, one might expect a change in the mass
distribution of white dwarfs in the LMC compared to other metallicity
environments. The mass distribution of the CSPNs should show the same
effect. The star formation history of the LMC should also be reflected in 
the average mass of CSPNs. We plan to compare the mass distribution 
of the CSPNs in two different metallicity environments: the LMC and the SMC
that are free of the biases that make the comparison with the Galactic CSs
very difficult.

We would like also to propose the LMC PNs as excellent tests probes to
study the gas-dynamic processes and wind injection rates. By comparing the
correlations between the nebular radius and the stellar luminosity with
numerical models pursued for LMC metallicities we will be able to constrain
the the efficiency of the wind driven mechanisms during the post-AGB phase.
Finally we do not find any strong evidence of morphological segregation as a
function of the progenitor mass, although
this point will be explored further once we analyze a larger sample of 
objects already observed with the {\it HST}.

\acknowledgments 
We would like to thank Jes\'us Ma\'{\i}z-Apell\'aniz for sharing his
experience on photometric analysis. This work has been supported by NASA
through grants GO-08271.01-97A and GO-08702 from Space Telescope Science 
Institute, which is operated by the Association of Universities for Research 
in Astronomy.

\begin{deluxetable}{lccc}
\tabletypesize{\scriptsize}
\tablenum{1}
\tablewidth{0pt}
\tablecaption{OBSERVATIONS}
\tablehead{
\multicolumn{1}{c}{Name}& 
\multicolumn{1}{c}{Instrument/Configuration} & 
\multicolumn{1}{c}{Integration (s)} &
\multicolumn{1}{c}{CS Detection}} 
\startdata
J~41    & STIS/50CCD  & 300   &   YES \\
SMP~4   & STIS/50CCD  & 120   &   YES \\
SMP~9   & STIS/50CCD  & 120   &   NO  \\
        & WFPC2/F547M & 1600  &   NO  \\         
SMP~10  & STIS/50CCD  & 120   &   YES \\
SMP~13  & STIS/50CCD  & 120   &   YES \\
SMP~16  & STIS/50CCD  & 300   &   NO  \\
        & WFPC2/F547M & 1600  &   NO  \\
SMP~18  & STIS/50CCD  & 120   &   YES \\
SMP~19  & STIS/50CCD  & 120   &   YES \\
        & WFPC2/F547M & 1600  &   NO  \\
SMP~25  & STIS/50CCD  & 120   &   YES \\
SMP~27  & STIS/50CCD  & 120   &   YES \\
SMP~28  & STIS/50CCD  & 120   &   YES \\
SMP~30  & STIS/50CCD  & 120   &   YES \\
        & WFPC2/F547M & 1600  &   NO  \\
SMP~31  & STIS/50CCD  & 120   &   YES \\
SMP~33  & WFPC2/F547M & 1600  &   NO  \\
SMP~34  & STIS/50CCD  & 120   &   YES \\
SMP~42  & WFPC2/F547M & 1600  &   YES \\
SMP~46  & STIS/50CCD  & 120   &   NO  \\
        & WFPC2/F547M & 1600  &   NO  \\
SMP~50  & WFPC2/F547M & 1600  &   YES \\
SMP~52  & WFPC2/F547M & 1600  &   YES \\
SMP~53  & STIS/50CCD  & 120   &   NO  \\
        & WFPC2/F547M & 800   &   NO  \\
SMP~56  & WFPC2/F547M & 1600  &   YES \\
SMP~58  & STIS/50CCD  & 120   &   YES \\
SMP~59  & STIS/50CCD  & 120   &   YES \\
SMP~63  & WFPC2/F547M & 800   &   YES \\
SMP~65  & STIS/50CCD  & 120   &   YES \\
SMP~71  & STIS/50CCD  & 120   &   NO  \\
SMP~78  & STIS/50CCD  & 120   &   NO  \\
        & WFPC2/F547M & 800   &   NO  \\
SMP~79  & STIS/50CCD  & 120   &   NO  \\
SMP~80  & STIS/50CCD  & 120   &   YES \\ 
SMP~81  & STIS/50CCD  & 120   &   YES \\ 
SMP~93  & STIS/50CCD  & 120   &   NO  \\
SMP~94  & STIS/50CCD  & 120   &   YES \\ 
SMP~95  & STIS/50CCD  & 120   &   NO  \\
SMP~100 & STIS/50CCD  & 120   &   YES  \\
SMP~102 & STIS/50CCD  & 120   &   YES  \\
\enddata
\end{deluxetable}

\begin{deluxetable}{lclcc}
\tabletypesize{\scriptsize}
\tablenum{2}
\tablewidth{0pt}
\tablecaption{MAGNITUDES, EXTINCTION, AND \heii~FLUXES}
\tablehead{
\multicolumn{1}{c}{Name}& 
\multicolumn{1}{c}{V$\pm\sigma$} & 
\multicolumn{1}{c}{E$_{B-V}$} & 
\multicolumn{1}{c}{I(\heii)$\pm\sigma$} & 
\multicolumn{1}{c}{Reference}
} 
\startdata
J 41   &      19.88$~\pm~$0.07& 0.00 & 8$~\pm~$4 & BL \\
SMP 4  &      21.19$~\pm~$0.08& 0.09  & 38.10$~\pm~$3.81  & VDM  \\        
SMP 9  &$\ge$~22.18& 0.16   & 39.5$~\pm~$3.91   & VDM  \\ 
       &$\ge$~22.35&       &                   &      \\
SMP10  &      20.72$~\pm~$0.07& 0.11   &  5.0$~\pm~$1      & VDM   \\ 
SMP13  &      21.80$~\pm~$0.19& 0.06   & 42.17$~\pm~$2.1   & MED2  \\
SMP16  &$\ge$~22.15& 0.10   & 69.75$~\pm~$3.5   & MED2 \\
       &$\ge$~21.74\tablenotemark{a}&       &                        \\
SMP18  &      18.97$~\pm~$0.07& 0.05   &      0.0          & VDM \\     
SMP19  &      22.25$~\pm~$0.43& 0.13   &  45.15$~\pm~$2.3  & MED2 \\
       &$\ge$~20.38\tablenotemark{a}&       &                   &   \\
SMP 25 &      16.73\tablenotemark{b}$~\pm~$0.09& 0.09 &    0.0             & JK93 \\     
SMP 27 &      18.68$~\pm~$0.06& 0.04 &    0.0             & SHAW \\        
SMP 28 & \nodata                   &0.23 &             \nodata &   \\  
SMP 30 &      24.38$~\pm~$0.28& 0.08  &          \nodata  &   \nodata \\     
       &$\ge$~23.67\tablenotemark{a}  &       &                   &   \\ 
SMP 31 &      17.04$~\pm~$0.05& 0.38  &   0.0             & MED \\     
SMP 33 &$\ge$~20.08\tablenotemark{a}  & 0.26  &   44.8 $~\pm~$4   & SHAW\\    
SMP 34 &      17.93$~\pm~$0.08& 0.04  &   20.6$~\pm~$4   & PALEN  \\    
SMP 42 &      19.03\tablenotemark{a}$~\pm~$0.02& 0.16  &    7.8$~\pm~$0.8 & MED2\\     
SMP 46 &$\ge$~21.42& 0.13  &   31.5$~\pm~$1.6& MED2   \\   
       &$\ge$~21.00\tablenotemark{a}&       &                   &      \\
SMP 50 &      18.89\tablenotemark{a}$~\pm~$0.07& 0.13  &   19.0$~\pm~$1& MED  \\    
SMP 52 &      19.84\tablenotemark{a}$~\pm~$0.16& 0.20  &   25.0$~\pm~$1.25& MED\\   
SMP 53 &$\ge$~18.42& 0.09  &   0.0            & MBC \\
       &$\ge$~18.01\tablenotemark{a}&       &                   & MBC \\
SMP 56 &      17.79\tablenotemark{a}$~\pm~$0.01& 0.08  &   0.4 $~\pm~$0.08 & SHAW   \\ 
SMP 58 &    \nodata &0.08  &   1.8$~\pm~$0.18 & MED     \\     
SMP 59 &      20.10$~\pm~$0.06& 0.00  &   80.9$~\pm~$8.09 & SHAW     \\ 
SMP 63 &      17.58\tablenotemark{a}$~\pm~$0.04& 0.16  &   0.0            & MED \\  
SMP 65 &      18.11$~\pm~$0.05& 0.16  &   0.0            & MED  \\      
SMP 71 &$\ge$~18.67& 0.17  &   39.9$~\pm~$4& MBC \\      
SMP 78 &$\ge$~17.34\tablenotemark{b}& 0.15  &   29.6$~\pm~$1.5& MED2\\
       &$\ge$~17.54\tablenotemark{a}&       &        &        \\
SMP 79 &$\ge$~17.23& 0.13  &    \nodata       &        \\
SMP 80 &      18.24$~\pm~$0.09& 0.06  &   0.0           &  PALEN \\ 
SMP 81 &      16.38\tablenotemark{b}$~\pm~$0.05& 0.17  &   0.0           &  MBC \\ 
SMP 93 &$\ge$~25.55& 0.00  &   52.4$~\pm~$5.2& MBC  \\     
SMP 94\tablenotemark{c} & 15.22\tablenotemark{b}$~\pm~$0.04& 0.74 &54.7$~\pm~$2.7 &  MED  \\
SMP 95 &$\ge$~23.40& 0.08  &   28.6$~\pm~$4.3 &  VDM   \\    
SMP 100&      21.84$~\pm~$0.19& 0.014 &   38.7$~\pm~$3.9 &  SHAW  \\  
SMP 102&      22.15$~\pm~$0.19& 0.00  &   75.4$~\pm~$7.5 &  SHAW   \\
\enddata
\tablecomments{The $\ge$ symbol refers to lower limit to the magnitude when
  the CS is not detected} 
\tablenotetext{a}{WFPC2 data} 
\tablenotetext{b}{CS saturated}
\tablenotetext{c}{Probably not a PN.}
\tablerefs{(BL)\cite{BL}; (MED2) \cite{MED2}; (VDM) \cite{VDM}; (JK93)
  \cite{JK93}; (SHAW) Shaw et al. 2003 in preparation; (MBC)
  \cite{MBC};(PALEN) Palen et al. 2003 in preparation.} 
\end{deluxetable}

\begin{deluxetable}{lccccccc}
\tabletypesize{\scriptsize}
\tablenum{3}
\tablewidth{0pt}
\tablecaption{CS PARAMETERS}
\tablehead{
\multicolumn{1}{c}{Name}& 
\multicolumn{1}{c}{T$_{\rm eff}$(He~{\sc II})$\pm \sigma$} & 
\multicolumn{1}{c}{T$_{\rm eff}$(H)$\pm \sigma$}&
\multicolumn{1}{c}{$\log L_Z/L_{\sun}$$\pm \sigma$} & 
\multicolumn{1}{c}{$\log L_Z/L_{\sun}$$\pm \sigma$}& 
\multicolumn{1}{c}{M$_{\rm V}\pm \sigma$}&
\multicolumn{1}{c}{$\log L_*/L_{\sun}$$\pm \sigma$}&
\multicolumn{1}{c}{BC$\pm \sigma$}\\
\multicolumn{1}{c}{}& 
\multicolumn{1}{c}{10$^3$  {\rm K}} & 
\multicolumn{1}{c}{10$^3$  {\rm K}}&
\multicolumn{1}{c}{(\heii)} & 
\multicolumn{1}{c}{ (H) }& 
\multicolumn{1}{c}{}&
\multicolumn{1}{c}{}&
\multicolumn{1}{c}{}
}
\startdata
J~41   &   60.1$\pm$ 3.5 &   30.4$\pm$ 1.8 &         3.24$\pm$0.07 &         2.47$\pm$0.06 &          1.36$\pm$0.07 &       3.37$\pm$0.07 &  -5.03$\pm$0.17  \\
SMP~4  &   89.9$\pm$ 7.2 &   48.1$\pm$ 7.4 &         3.31$\pm$0.09 &         2.56$\pm$0.17 &          2.67$\pm$0.08 &       3.32$\pm$0.10 &  -6.23$\pm$0.24  \\
SMP~9  &  114.0$\pm$10.2 &   77.6$\pm$13.7 &   $\le$~3.29          &  $\le$~2.89           &  $\ge$~3.66 & $\le$~3.25          &  -6.93$\pm$0.27  \\
SMP~10 &   74.6$\pm$ 4.3 &   57.1$\pm$ 7.9 &         3.31$\pm$0.07 &         2.98$\pm$0.15 &          2.20$\pm$0.07 &       3.29$\pm$0.07 &  -5.67$\pm$0.17  \\
SMP~13 &  129.1$\pm$11.9 &   99.1$\pm$18.2 &         3.49$\pm$0.11 &         3.16$\pm$0.20 &          3.28$\pm$0.19 &       3.51$\pm$0.13 &  -7.30$\pm$0.27  \\
SMP~16 &  141.9$\pm$17.3 &   97.5$\pm$21.8 &   $\le$~3.45          &  $\le$~3.07          &  $\ge$~3.62 & $\le$~3.54          &  -7.58$\pm$0.36  \\
SMP~18 &      \nodata    &   30.9$\pm$ 2.8 &             \nodata   &         2.91$\pm$0.09 &          0.45$\pm$0.07 &       2.94$\pm$0.11 &  -3.05$\pm$0.27  \\
SMP~19 &  143.5$\pm$17.5 &  119.1$\pm$27.9 &         3.52$\pm$0.12 &         3.29$\pm$0.21 &          3.73$\pm$0.43 &       3.45$\pm$0.23 &  -7.61$\pm$0.36  \\
SMP~25\tablenotemark{a} &      \nodata    &   33.7$\pm$ 2.6 &          \nodata             &          3.95$\pm$0.08 &  -1.79$\pm$0.09 &   3.94$\pm$0.10 &  -3.31$\pm$0.23  \\
SMP~27 &      \nodata    &   28.3$\pm$ 2.9 &             \nodata   &         2.92$\pm$0.10 &          0.16$\pm$0.06 &       2.95$\pm$0.13 &  -2.79$\pm$0.31  \\
SMP~28\tablenotemark{b} &      \nodata    &   17.1$\pm$ 1.1 &             \nodata          &         4.14$\pm$0.05 &     \nodata&  \nodata       &   \nodata        \\
SMP~30 &      \nodata    &  149.3$\pm$37.8 &             \nodata   &         2.66$\pm$0.27 &          5.86$\pm$0.28 &       2.65$\pm$0.32 &  -7.73$\pm$0.75  \\
SMP~31 &      \nodata    &   28.6$\pm$ 2.4 &             \nodata   &         4.01$\pm$0.08 &         -1.48$\pm$0.05 &       3.62$\pm$0.10 &  -2.82$\pm$0.25  \\
SMP~33 &  110.9$\pm$11.2 &   69.7$\pm$13.8 &   $\le$~4.16          &  $\le$~3.64          &  $\ge$~1.55 & $\le$~4.06          &  -6.85$\pm$0.30  \\
SMP~34 &   67.8$\pm$ 3.6 &   32.0$\pm$ 2.9 &         4.22$\pm$0.06 &         3.35$\pm$0.10 &         -0.59$\pm$0.08 &       4.29$\pm$0.07 &  -5.38$\pm$0.16  \\
SMP~42 &   66.4$\pm$ 0.8 &   37.7$\pm$ 0.3 &         3.89$\pm$0.02 &         3.23$\pm$0.01 &          0.51$\pm$0.02 &       3.82$\pm$0.02 &  -5.32$\pm$0.03  \\
SMP~46 &  119.4$\pm$35.4 &   93.8$\pm$55.4 &   $\le$~3.20          &  $\le$~2.84          &  $\ge$~2.90 & $\le$~3.65          &  -7.07$\pm$0.88  \\
SMP~50 &   80.4$\pm$ 0.9 &   46.0$\pm$ 0.8 &         4.15$\pm$0.02 &         3.49$\pm$0.01 &          0.37$\pm$0.07 &       4.11$\pm$0.03 &  -5.89$\pm$0.03  \\
SMP~52 &  100.9$\pm$ 2.6 &   69.8$\pm$ 3.5 &         4.13$\pm$0.04 &         3.68$\pm$0.01 &          1.32$\pm$0.16 &       4.00$\pm$0.07 &  -6.57$\pm$0.08  \\
SMP~53 &      \nodata    &   43.2$\pm$ 4.1 &     \nodata           &  $\le$~3.60          &  $\ge$~-0.1     & $\le$~3.63          &  -4.04$\pm$0.28  \\
SMP~56 &   45.9$\pm$ 2.0 &   29.0$\pm$ 3.1 &         3.85$\pm$0.05 &         3.35$\pm$0.10 &         -0.74$\pm$0.02 &       3.89$\pm$0.05 &  -4.23$\pm$0.13  \\
SMP~58\tablenotemark{b}  &   71.4$\pm$ 2.8 &   71.4$\pm$8.4       &3.49$\pm$0.05 &3.49$\pm$0.13 &         \nodata  &        \nodata     &    \nodata       \\
SMP~59 &   98.2$\pm$ 6.8 &   46.2$\pm$ 5.6 &         3.75$\pm$0.08 &         2.85$\pm$0.13 &          1.58$\pm$0.06 &       3.86$\pm$0.09 &  -6.49$\pm$0.21  \\
SMP~63 &      \nodata    &   38.8$\pm$ 0.4 &             \nodata   &         3.85$\pm$0.01 &         -0.94$\pm$0.04 &       3.77$\pm$0.02 &  -3.73$\pm$0.03  \\
SMP~65 &      \nodata    &   27.0$\pm$ 2.2 &             \nodata   &         3.25$\pm$0.08 &         -0.41$\pm$0.06 &       3.12$\pm$0.10 &  -2.65$\pm$0.24  \\
SMP~71 &   83.4$\pm$ 5.2 &   40.8$\pm$ 4.8 &   $\le$~4.34          &  $\le$~3.53          &  $\ge$~0.15             & $\le$~4.27 &  -6.00$\pm$0.18  \\
SMP~78 &   75.7$\pm$ 3.1 &   36.4$\pm$ 2.8 &   $\le$~4.67          &  $\le$~3.82          &  $\ge$~-0.99            & $\le$~4.61 &  -5.71$\pm$0.12  \\
SMP~79 &      \nodata    &   33.5$\pm$ 2.6 &     \nodata           &  $\le$~3.80          &  $\ge$~-1.30           & $\le$~3.79 &  -3.29$\pm$0.23  \\
SMP~80 &      \nodata    &   30.7$\pm$ 2.8 &             \nodata   &         3.21$\pm$0.09 &         -0.28$\pm$0.10 &       3.22$\pm$0.12 &  -3.03$\pm$0.27  \\
SMP~81\tablenotemark{a} &      \nodata    &   28.1$\pm$ 1.7  &             \nodata          &    3.99$\pm$0.07 & -2.14$\pm$0.06   & 3.86$\pm$0.08 &  -2.77$\pm$0.18  \\
SMP~93 &  372.4$\pm$97.6 &  526.0$\pm$158.5&   $\le$~3.25          &  $\le$~3.73           &  $\ge$~7.03& $\le$~3.04$\pm$0.39 & -10.45$\pm$0.78  \\
SMP~94\tablenotemark{a} &   59.3$\pm$ 2.6 &   21.1$\pm$ 1.4 &         6.00$\pm$0.05        &   4.88$\pm$0.06 & -3.30$\pm$0.05    & 5.21$\pm$0.06 &  -4.99$\pm$0.13  \\
SMP~95 &  146.2$\pm$28.6 &  152.4$\pm$57.1 &   $\le$~2.81          &  $\le$~2.93          &  $\ge$~4.88 & $\le$~3.08 &  -7.67$\pm$0.58  \\
SMP~100&  127.1$\pm$11.7 &   99.5$\pm$18.3 &         3.39$\pm$0.11 &         3.09$\pm$0.20 &          3.31$\pm$0.19 &       3.47$\pm$0.13 &  -7.25$\pm$0.27  \\
SMP~102&  131.8$\pm$12.4 &   82.4$\pm$14.2 &         3.30$\pm$0.11 &         2.72$\pm$0.18 &          3.63$\pm$0.19 &       3.39$\pm$0.14 &  -7.36$\pm$0.28  \\
\enddata
\tablecomments{The $\ge$ symbol refers to lower limit to the magnitude when
  the CS is not detected, luminosities for those cases are therefore upper
  limits and are preceded by a $\le$ symbol.} 
\tablenotetext{a}{The photometry was performed on saturated data}
\tablenotetext{b}{Temperature and luminosity for this nebula were derived
  from crossover analysis. The temperatures should be considered upper
  limits.}  
\end{deluxetable}

\begin{deluxetable}{lccl}
\tabletypesize{\scriptsize}
\tablenum{4}
\tablewidth{0pt}
\tablecaption{STELLAR MASSES}
\tablehead{
\multicolumn{1}{c}{Name}&
\multicolumn{1}{c}{Morphology}& 
\multicolumn{1}{c}{M [\Mso]} & 
\multicolumn{1}{l}{Comments}}
\startdata
J~41     & E(bc)&  0.59 & Interpolation of He-burning tracks \\
SMP~4    & E    &  0.58 & He-burning track \\
SMP~10   & P    &  0.58 & He-burning track \\
SMP~13   & R(bc)&  0.63 & He-burning track \\  
SMP~19   & E(bc)&  0.63 & He-burning track  \\
SMP~30   & B    &  0.67\tablenotemark{a} & He-burning track \\
         &      &  0.67\tablenotemark{a} & H-burning track \\
SMP~31   & R    &  0.59\tablenotemark{a} & Interpolation of He-burning track\\
SMP~34   & E    &  0.84 & L-core mass relation H-burning tracks \\
SMP~42   & Q    &  0.67 & Interpolation of He-burning tracks \\
SMP~50   & E(bc)&  0.75 & Interpolation of H-burning tracks \\
SMP~52   & R(bc)&  0.70 & Interpolation of H-burning tracks \\
SMP~56   & R    &  0.68 & He-burning track  \\
SMP~59   & Q    &  0.65 & Extrapolation of H-burning tracks\\
         &      &  0.69 & Extrapolation of He-burning tracks   \\
SMP~63   & R   &  0.64\tablenotemark{a} & Interpolation of He-burning tracks\\
SMP~100  & Q&  0.63 & He-burning track  \\
SMP~102  & E(Bc) &  0.60 & Extrapolation of He-burning tracks  \\
\enddata
\tablenotetext{a}{Derived from Hydrogen Zanstra analysis and therefore rather
  uncertain (see text). Note that the masses derived from He-burning tracks
  might be slightly smaller if they were derived from H-burning tracks.} 
\end{deluxetable}

\begin{figure}
\plotone{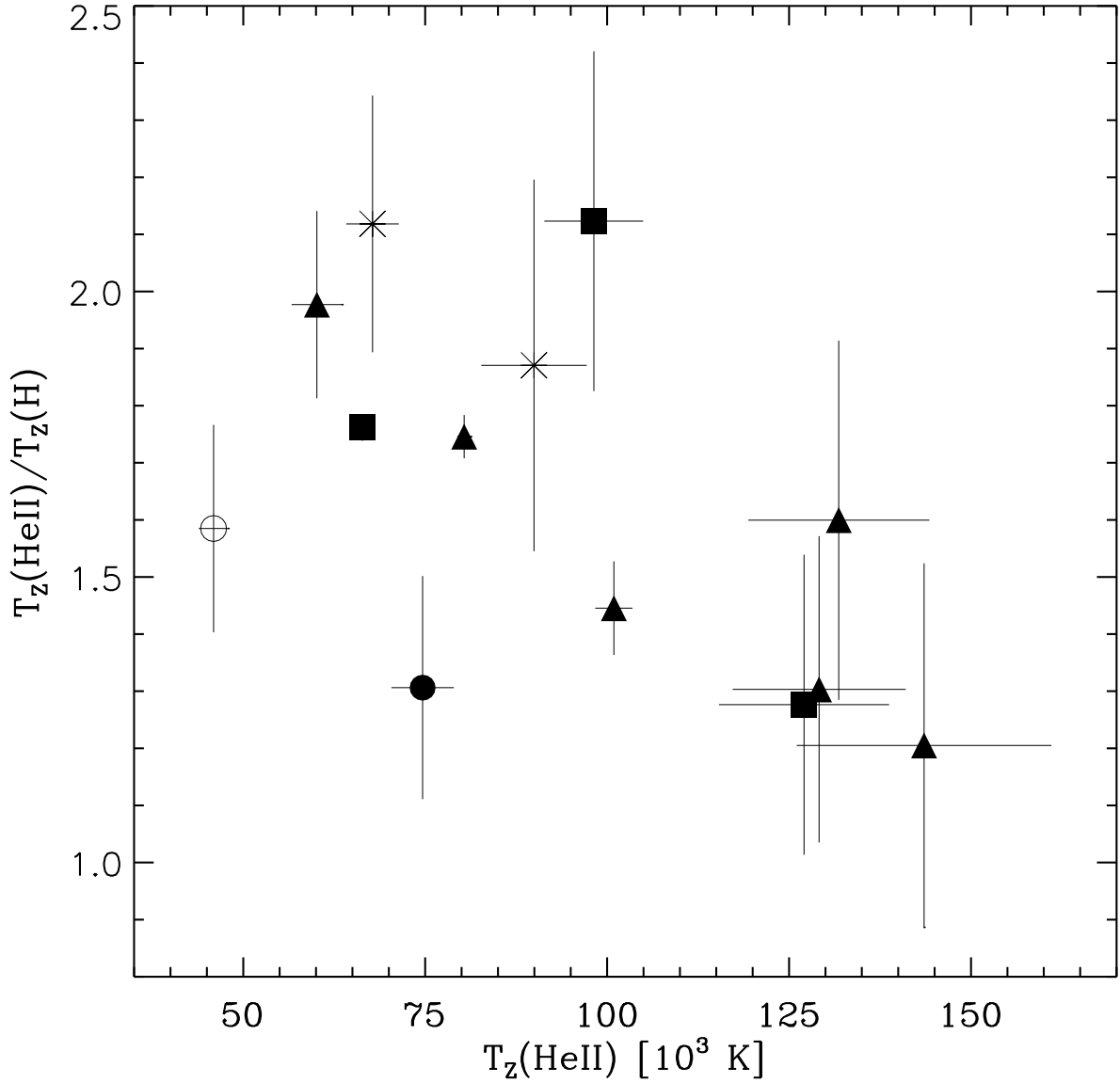}
\caption[ ]{Ratio of Zanstra temperatures versus the He~{\sc ii} Zanstra
  temperature. The symbols represent the morphological types of the hosting
  nebulae: round (open
  circles), elliptical (asterisks), bipolar and quadrupolar (squares),
  bipolar core (triangles) and point-symmetric (filled circles).
\label{f1.eps}}
\end{figure}

\begin{figure}
\plotone{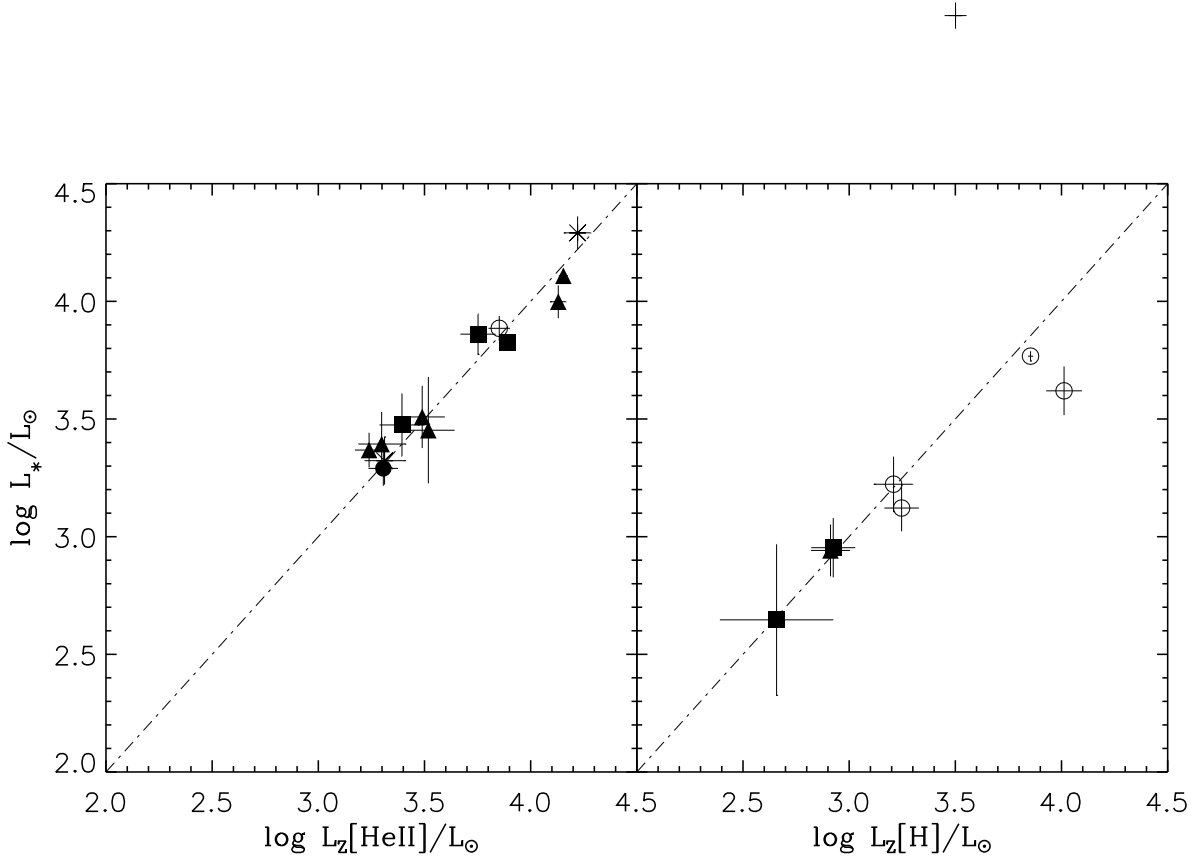}
\caption[ ]{Left panel: Stellar luminosity derived from the observed
  magnitude (in solar units logarithm scale) against
  Zanstra luminosity 
  from \heii~temperature (T$_Z$[\heii]). Right panel: the same but against
  the Zanstra luminosity derived from hydrogen temperature (T$_Z$[H]). The
  symbols represent the morphology of the hosting nebular and are as in
  Fig.~1. The dotted lines represents the 1:1 relation .
\label{f2.eps}}
\end{figure}

\begin{figure}
\plotone{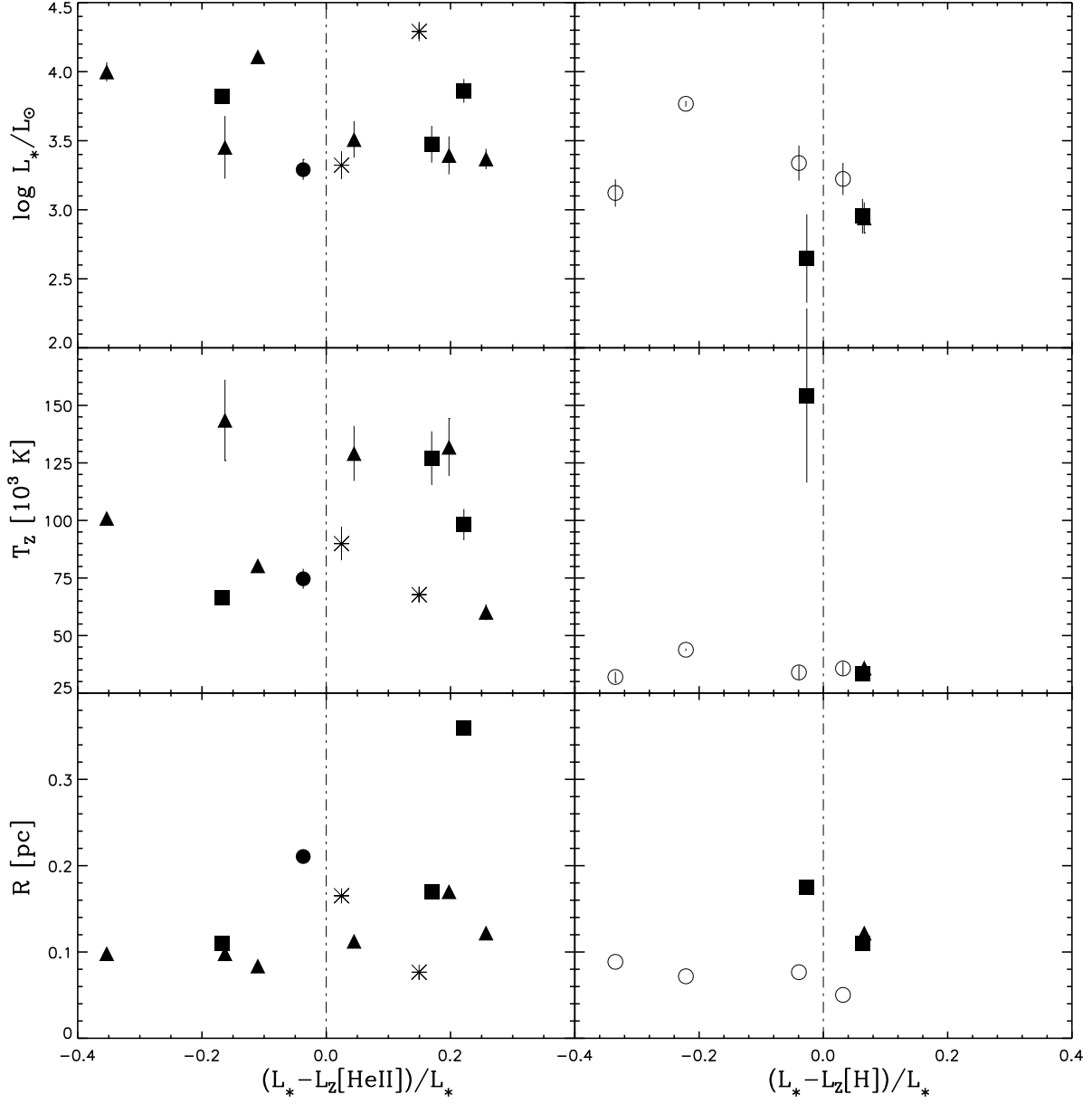}
\caption[ ]{Relative differences between the observed, L$_*$, and derived
  Zanstra Luminosities 
versus $\log$~L$_*$/L$_\odot$, Zanstra temperature, and nebular photometric
radius. The left and right panels are for the Luminosities derived from the
\heii~and H~{\sc i} Zanstra Temperatures respectively. The symbols are as in
Fig.~1 and the doted lines represent equal luminosity. 
\label{f3.eps}}
\end{figure}

\begin{figure}
\plotone{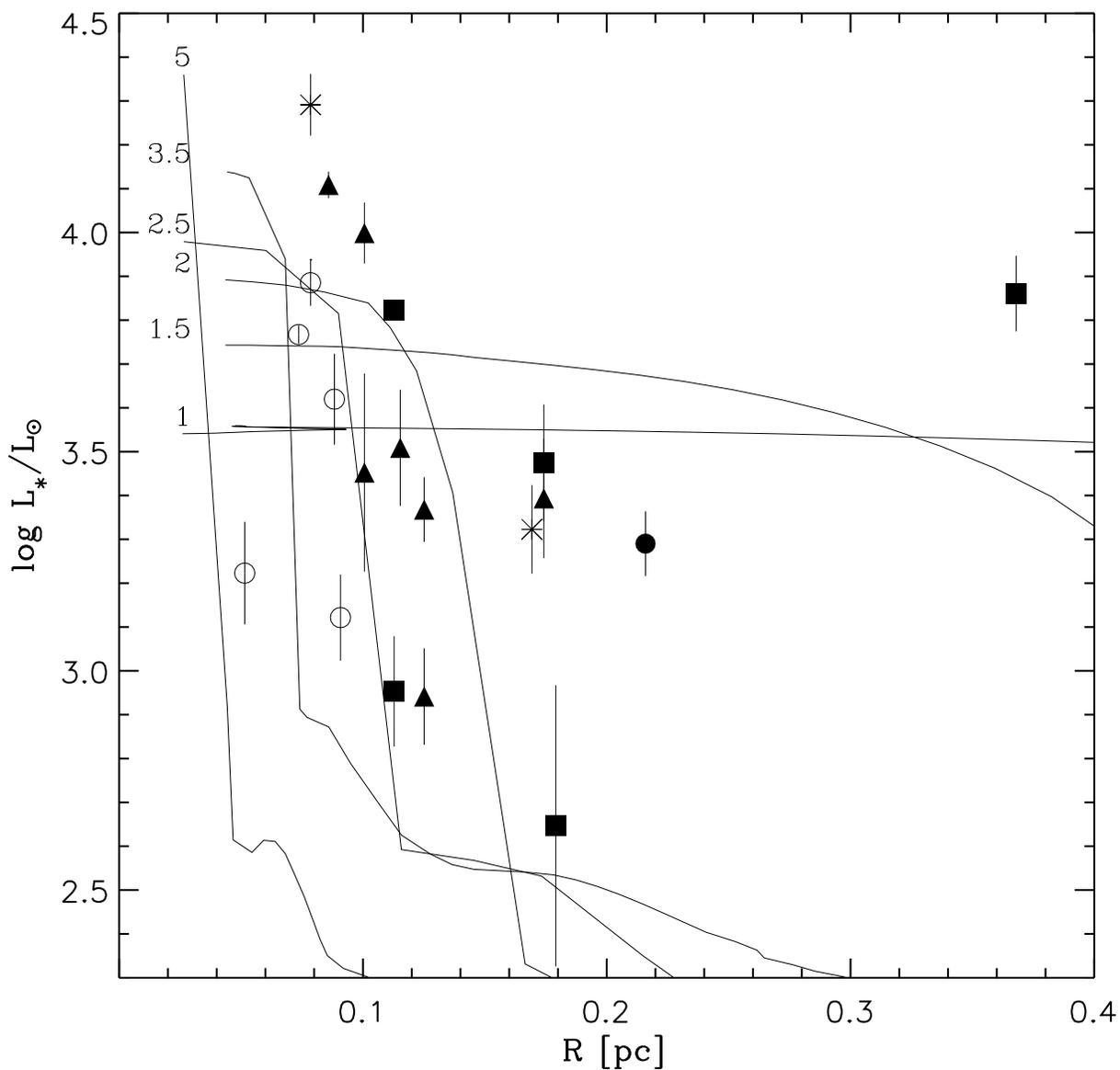}
\caption[ ]{The points represent the logarithm of the observed luminosity
  versus the physical radius 
  of the nebulae. The symbols are as in Fig.~1. The solid lines represent the
  evolution of the nebular radius versus the stellar luminosity taken from the
  numerical simulations of \cite{Vmg:02} for Galactic PN, each line has been
  marked with the initial mass of the progenitor used in the hydrodynamical
  simulations.
\label{f4.eps}}
\end{figure}

\begin{figure}
\plotone{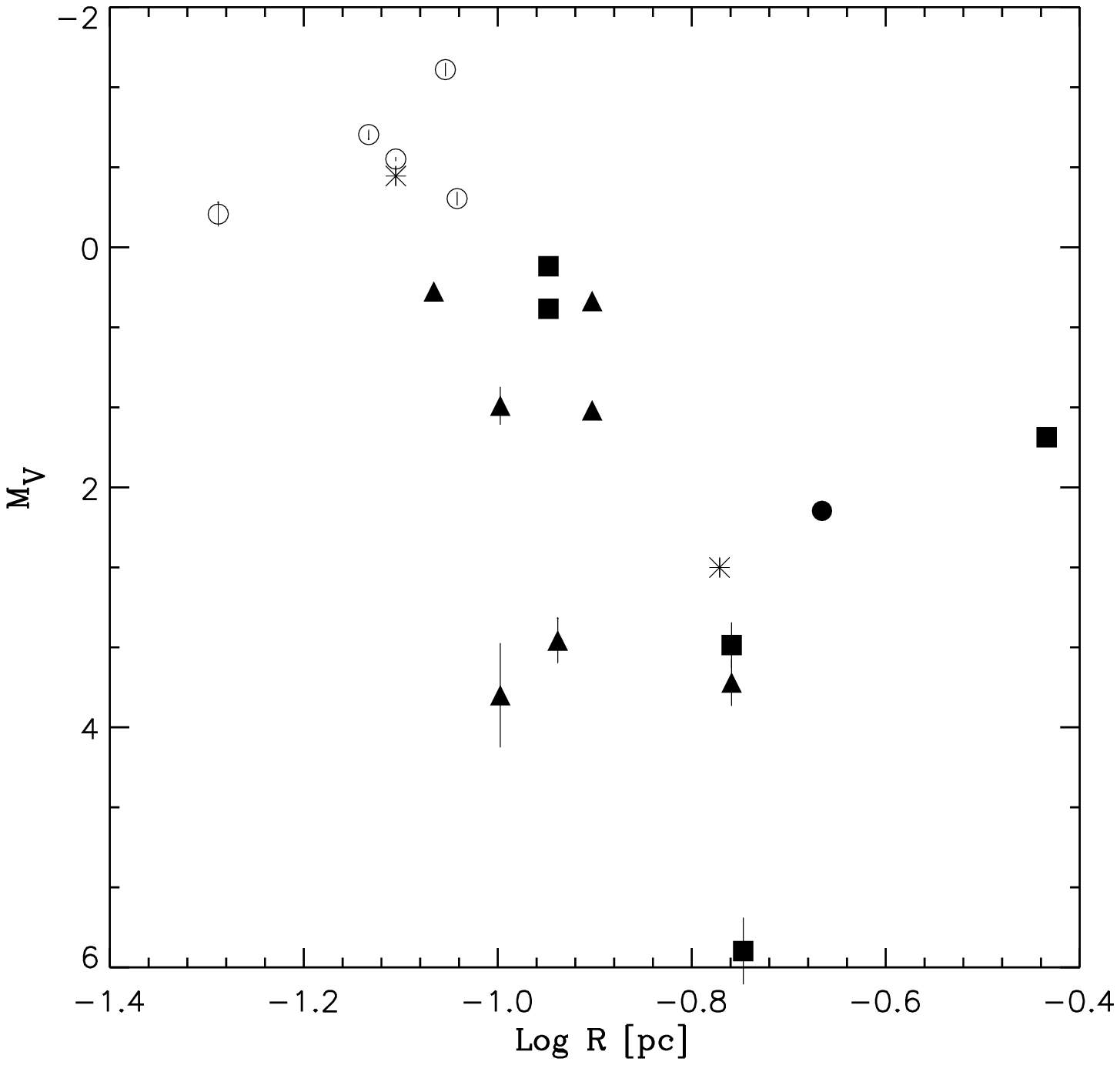}
\caption[ ]{Absolute visual magnitude
  versus the physical radius 
  of the nebulae. The symbols are as in Fig.~1. 
\label{f5.eps}}
\end{figure}

\begin{figure}
\plotone{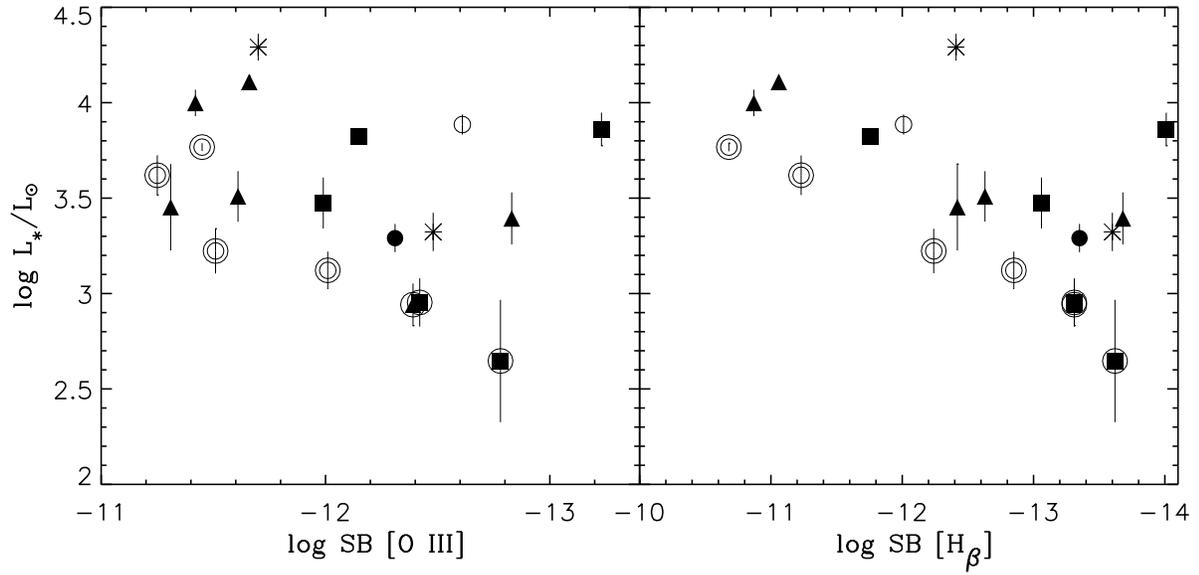}
\caption[ ]{Logarithm of the observed luminosity
  versus the surface brightness of the nebula in the \oiii~and \hb~lines. The
  symbols are as in Fig.~1, the objects with hydrogen Zanstra temperatures
  have been surrounded with a circle.
\label{f6.eps}}
\end{figure}

\begin{figure}
\plotone{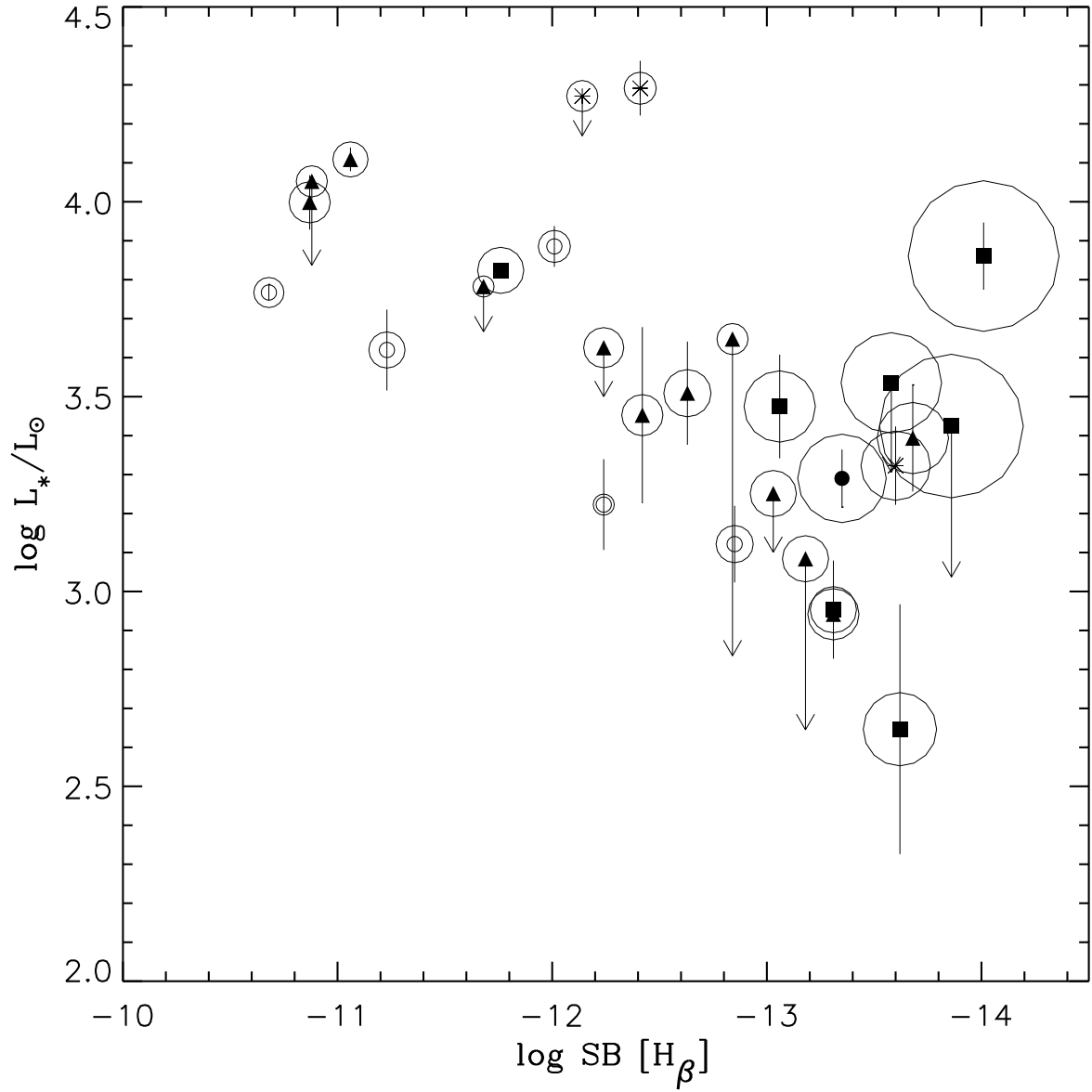}
\caption[ ]{Same as Fig.~6 but only for the \hb~line. We have also plotted 
  the upper limits of the Luminosity for those objects where the CS was not
  detected. The arrows mean that the magnitude was a lower limit and so the
  luminosity is an upper limit. Each data point is surrounded by a circle
  proportional to the radius of the hosting nebula. 
\label{f7.eps}}
\end{figure}

\begin{figure}
\plotone{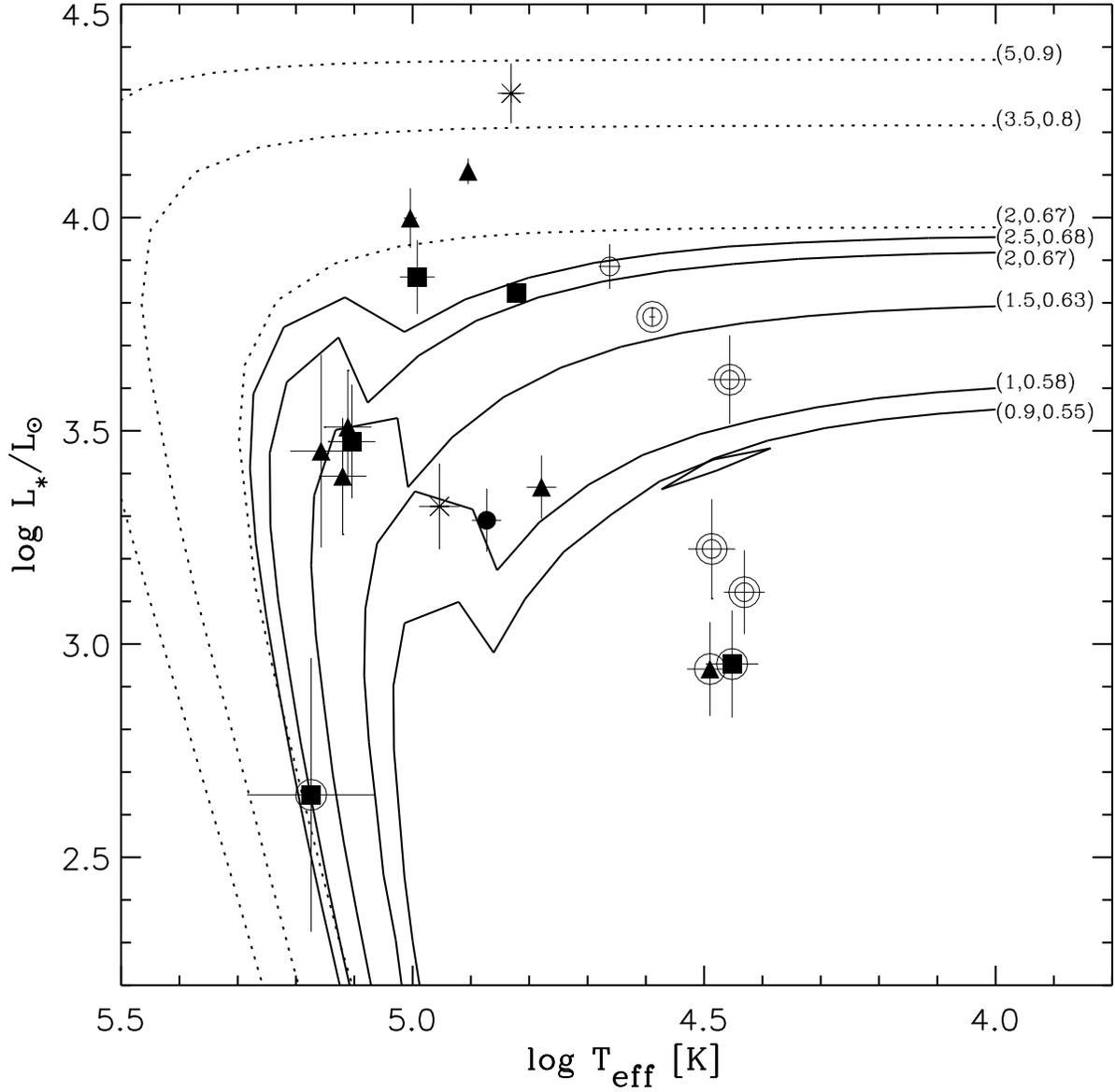}
\caption[ ]{HR diagram for the CSPNs. Symbols as in Fig.~1. We have
  surrounded by a circle those points for which the H I Zanstra temperatures 
  has been used. Evolutionary tracks are for LMC metallicities from
  Vassiliadis \& Wood (1994), the initial and core masses are marked on
  each track. The solid lines are for He-burners and the dotted lines for
  hydrogen burners.
\label{f8.eps}}
\end{figure}

\end{document}